# Quantum Metrology in Transmission Electron Microscopy

**Shiran Even-Haim*[1], Ethan Nussinson*[1], Roni Ben-Maimon[2], Alexey Gorlach[1], Ron Ruimy[1], Ephraim Shahmoon[2], Osip Schwartz†[3], Ido Kaminer†[1]**

[1]*Faculty of Electrical & Computer Engineering and Solid-State Institute, Technion - Israel Institute of Technology, Haifa 3200003, Israel*

[2]*Department of Chemical & Biological Physics, Weizmann Institute of Science, Rehovot 7610001, Israel*

[3]*Department of Physics of Complex Systems, Weizmann Institute of Science, Rehovot 7610001, Israel*

*†*equal contributors*

†*kaminer@technion.ac.il* †*osip@weizmann.ac.il*

## Abstract

**Quantum metrology experiments in atomic physics and quantum optics have demonstrated measurement accuracy beyond the shot-noise limit via multi-particle entanglement. At the same time, electron microscopy—an essential tool for high-resolution imaging across many fields of science and technology from materials to life sciences—is severely constrained in its signal-to-noise ratio (SNR) by shot noise, due to the dose limit imposed by electron beam-induced damage. Here, we show theoretically that quantum metrology methods based on electron entanglement are a natural fit for improving the SNR in electron microscopy. We investigate the generation of the necessary entangled states through electron-electron Coulomb interactions and quantum non-demolition measurements. Our results connect the fields of quantum metrology and electron interferometry, paving the way toward electron microscopy with SNR beyond the shot-noise limit.**

## Introduction

Transmission electron microscopes (TEM) leverage the short de Broglie wavelength of electrons to achieve atomic resolution. However, electron beam-induced damage significantly restricts the TEM's ability to image many soft-matter samples, including most samples of interest in life sciences. For example, biological samples imaged in high-resolution cryogenic electron microscopy (cryo-EM) are constrained to an electron dose of around 20 electrons per square Angstrom [1] by radiation damage. At such low doses, shot noise severely limits the signal-to-noise ratio of cryo-EM images.

Cryo-EM has achieved extraordinary success in determining the structure of biological macromolecules with atomic resolution, overcoming the limited signal-to-noise ratio (SNR) by combining TEM image data from thousands or even millions of identical macromolecules or molecular complexes [2]. However, this approach cannot be applied to samples with unique, non-repeating structures, such as lamellae (slices) ion-milled from whole cells, often studied with cryogenic electron tomography (cryo-ET). As life science studies using electron microscopy increasingly focus on such samples, improving the SNR of cryo-EM images beyond the shot-noise limit becomes a key challenge.

To formally define shot noise, we note that in cryo-EM, samples are typically weak phase objects, imprinting a small position-dependent phase shift $\phi \ll 1$ on the electron wave function. In addition to these small spatial variations of the phase, the sample also introduces an overall phase offset that is not necessarily small; however, it can be disregarded here without loss of generality. Under ideal measurement conditions (perfect phase contrast), the phase measurement accuracy is limited by shot noise, resulting in a phase uncertainty of $\Delta\phi_{\mathrm{SQL}} = \frac{1}{2}N^{-1/2}$, where $N$ is the number of electrons per pixel, representing a fundamental constraint known as the standard quantum limit (SQL) [3–5]. For a typical pixel size of 1 Å and a dose limit of 20 e/Å$^2$, the shot-noise-limited phase measurement error is $\Delta\phi_{\mathrm{SQL}} \approx 0.11$ radians. Thus, in high-resolution cryo-EM images, the per-pixel SNR defined as $|\phi|/\Delta\phi$ is typically of order unity or less.

In practice, cryo-EM already achieves phase measurement accuracy close to the SQL. The sample-induced phase profile is conventionally converted to amplitude variation by defocusing the TEM's imaging system, which results in a contrast transfer function that oscillates with spatial frequency, achieving near-SQL SNR at the maxima of the oscillations. Recently developed Zernike-like phase plates for TEM [6,7] enable efficient conversion of the phase profile to amplitude modulation, approaching the SQL with a more uniform contrast transfer function. The electrons' spatial distribution in the image plane of the TEM is then detected by direct electron detectors [8,9], which enable near-perfect counting of electrons passing through each pixel. Thus, further SNR improvement beyond the current state of the art calls for new methods with phase accuracy surpassing the SQL.

Quantum physics can be harnessed to improve the SNR beyond the SQL, with the ultimate bound set by the Heisenberg limit [3], corresponding to a phase uncertainty of $\Delta\phi_{\mathrm{H}} = N^{-1}$. Several quantum metrology approaches have been proposed for achieving sub-shot-noise-limited SNRs in electron microscopy [10–12]. These can be broadly grouped into two classes. One consists of single-electron signal enhancement approaches, such as multi-pass microscopy [13,14] or interaction-free measurements [10,15]. The other is based on multi-electron entangled states [16–18]. A pioneering effort in this direction explores using superconducting quantum circuits coupled to a TEM [11,19,20]. None of these approaches have yet achieved an experimental demonstration of quantum-enhanced electron interferometry.

At the same time, in other fields, entanglement-based quantum metrology has achieved measurement accuracy well beyond the shot-noise limit. A central concept within the field of quantum metrology is quantum *squeezing*, the reduction of the quantum fluctuations of an observable via entanglement to improve measurement accuracy. Metrological improvements via squeezing were demonstrated in, e.g., light interferometry [21–26], nonlinear microscopy [27–29], or dark matter searches [30]. However, unlike optical photons where squeezing has been experimentally demonstrated for decades [31], extending these ideas of squeezing remains a major pursuit for entire communities of energetic particles such as free electrons [12,32,33] and high-energy photons [34,35] (EUV or X-rays). A particular type of squeezing, known as *spin squeezing* [36], applicable to ensembles of two-level quantum systems, plays a central role in quantum metrology experiments with atomic ensembles [37–45].

In this work, we show that entanglement-based quantum metrology enables a significant improvement of phase measurement accuracy under realistic parameters. We envision an electron microscope implemented as a scanning interferometer [46,47] that creates images pixel by pixel by measuring the sample-induced phase shift at each point (Fig. 1a, b). Each electron passing through the interferometer is in a quantum superposition of being either in one arm of the interferometer or in the other. We use the two possible paths as basis states, analogous to the two states of spin-½ particles, to define a pseudospin degree of freedom (unrelated to the physical spin) for each electron [36]. The pseudospin representation of electron paths in an interferometer enables the application of the spin squeezing formalism to electron microscopy.

We then explore practical means to generate multi-electron quantum states that are spin-squeezed with respect to this pseudospin degree of freedom and show that such states enable a substantial reduction of shot noise-induced phase uncertainty under conditions relevant to cryo-EM. Spin squeezing of free electrons thus offers a natural avenue toward quantum-enhanced electron microscopy with SNR well above the SQL.

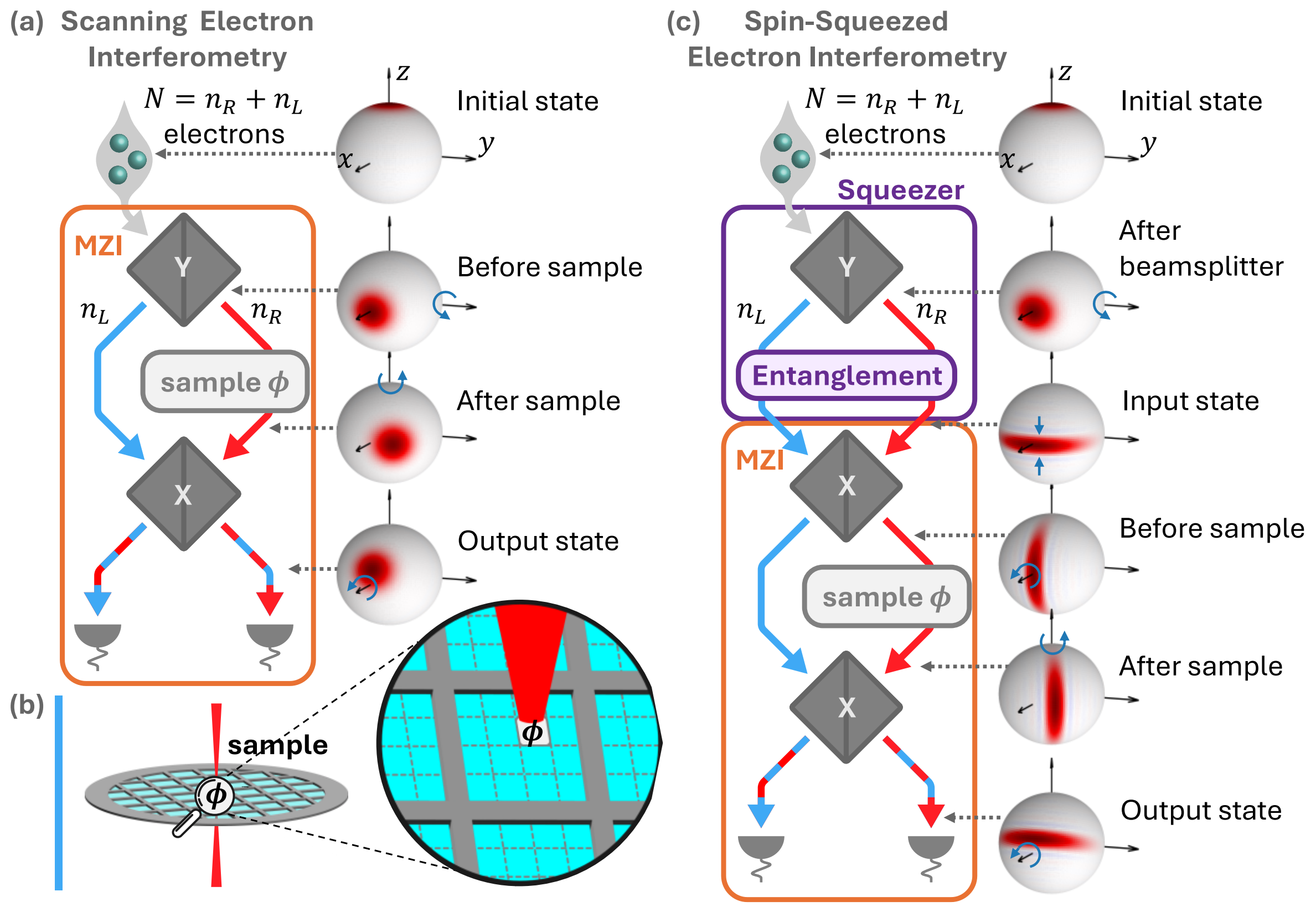

**Fig. 1: Squeezing in electron interferometry. (a)** Scanning electron Mach-Zehnder interferometer (MZI): multi-electron states are shown on the Bloch sphere, representing Wigner functions at different stages. An incoming $N$-electron state is rotated by $\pi/2$ about the $y$-axis by the first beamsplitter, initializing a superposition between the two arms. The electrons in the right arm interact with the sample, acquiring a phase $\phi$, corresponding to $\phi$-rotation around the $z$-axis. Finally, a second beamsplitter induces $\pi/2$ rotation about the $x$-axis, and a measurement of the number of electrons in each port is performed (measuring the $z$ projection of the collective pseudospin). **(b)** The sample is scanned (red beam) pixel by pixel. **(c)** Spin-squeezed electron interferometry: we introduce a preparation stage, referred to as the squeezer, where a multi-electron squeezed state is prepared before being coupled into the MZI. The first beamsplitter of the MZI aligns the squeezed axis with the direction of the phase $\phi$ accumulation, thus enhancing the precision of the phase measurement.

This paper is organized as follows. First, we introduce the pseudospin and spin squeezing in the context of electron interferometry. Then, we outline two different approaches for building an electron 'squeezer' device that prepares a batch of electrons in a pseudospin-squeezed state (Fig. 1c). The first is an interaction-based method, where the pseudospin-squeezed states are generated by electron-electron interactions that entangle the electrons. The second approach is a measurement-based method, where the pseudospin-squeezed states are generated by the electron interaction with a measuring device that implements a quantum non-demolition measurement that entangles the electrons. In both cases, we determine the upper bound of the achievable phase measurement accuracy based on quantum Fisher information and evaluate the phase accuracy attainable in a conventional interferometric measurement with such squeezed states. Furthermore, we outline potential experimental implementations for each method. We then turn to discussion and outlook in the Discussion section.

# Results

A Mach-Zehnder interferometer (Fig. 1a) measures the relative phase between two paths—one that interacts with the sample and the other that serves as a reference. We consider a scenario where a batch of electrons is used to measure the phase at each pixel of the sample. For each electron, we denote the electron states passing through the right and left arms of the interferometer as $|R\rangle_n$ and $|L\rangle_n$, respectively, where $n \in \{1, \dots, N\}$ denotes the index of an electron within the batch of $N$ electrons. In this notation, each electron is assigned a pseudospin operator described by the spin-½ Pauli operators $\hat{\sigma}_n^i$ with $i \in \{x, y, z\}$. The states $|L\rangle_n$ and $|R\rangle_n$ correspond to the north and south poles of the Bloch sphere representation of this pseudospin, respectively, and satisfy $\hat{\sigma}_n^z|L\rangle_n = |L\rangle_n$, $\hat{\sigma}_n^z|R\rangle_n = -|R\rangle_n$. A phase shift between the interferometer arms (e.g., due to interaction with a sample) corresponds to a rotation around the $z$-axis. A beamsplitter couples the two paths, performing rotations around the $x$ or $y$ axis, with the rotation angle set by the magnitude of its coupling coefficient.

We emphasize that the three-dimensional space introduced here does not represent the physical three-dimensional space. The rotation operators in this space should be understood as acting on the electrons' pseudospin degrees of freedom, i.e., as either coupling the two arms of the interferometer or shifting their relative phase.

## Electrons as quasi-distinguishable particles

An important consideration in electron microscopy is that the electron density is low enough such that the wave functions of different electrons in the beam do not overlap. As a result, exchange terms are negligible, and the antisymmetrized description gives the same physical predictions as a description in which the non-overlapping wave-packet modes serve as effective labels [48–50]. In this regime, the Pauli exclusion principle does not prohibit multiple electrons in the beam from traveling through the same arm of the interferometer. Thus, even though electrons are fundamentally indistinguishable particles, they can be treated as distinguishable under these conditions, allowing us to attribute individual path-pseudospin operators $\hat{\sigma}_n^i$ to each electron.

Furthermore, we assume that each electron's pseudospin wave function undergoes the same sequence of transformations in the interferometer. Under these conditions, the multi-electron wave function remains symmetric with respect to the exchange of pseudospins, and therefore an ensemble of $N$ electrons can be described as a single system with a spin $N/2$ [36,51]. The collective spin operators are given by $\hat{S}_i = \frac{1}{2}\sum_{n=1}^{N} \hat{\sigma}_n^i$. We define number operators in the left and right paths, $\hat{n}_L$ and $\hat{n}_R$, where $\hat{n}_R = \frac{N}{2} - \hat{S}_z$, such that the total number of electrons is conserved: $\hat{n}_L + \hat{n}_R = N$. The symmetric multi-electron state with $n_R$ electrons in the right path is denoted as $|n_L, n_R\rangle = |N - n_R, n_R\rangle$. These collective spin states are also known as Dicke states.

While collective path-pseudospin squeezing has not yet been demonstrated for free electrons, the collective-spin approach has been experimentally validated in several spin-squeezing platforms, including trapped ions [52], optical-lattice clocks [53], and programmable optical-tweezer arrays [54]. In these systems, as in our proposal, the particles occupy distinct, non-overlapping modes while their internal two-level degrees of freedom are treated collectively.

The interferometer operation is composed of a sequence of steps (Fig. 1a), each described as a rotation of the collective spin. In a conventional interferometer, with uncorrelated electrons, the input state is created by the first beamsplitter that applies a $y$-axis rotation operator $e^{\frac{i\pi}{2}\hat{S}_y}$ on the initial state $|N, 0\rangle$. This operator prepares each electron in the same state $\frac{1}{\sqrt{2}}(|L\rangle_n + |R\rangle_n)$, corresponding to an equal superposition of passing through the left and right arms. The multi-electron wavefunction then takes the form of the coherent spin state $|\psi_0\rangle = 2^{-N/2}\prod_{n=1}^{N}(|L\rangle_n + |R\rangle_n)$. The electrons then accumulate a relative phase between the two paths due to their interaction with the sample in the right arm of the

interferometer (Fig. 1b), represented by the operator $e^{i\phi\hat{S}_z}$, corresponding to a rotation around the $z$-axis by an angle $\phi$. A second beamsplitter then applies an $x$-axis rotation $e^{\frac{i\pi}{2}\hat{S}_x}$.

Finally, the spin projection along the $z$-axis is measured by detecting the difference in the number of electrons arriving in the two output ports of the interferometer for each individual measurement: $S_z = \frac{n_L - n_R}{2}$. The corresponding phase $\phi$ is then determined using the relation:

$$S_z = \frac{N}{2}\sin\phi\,. \tag{1}$$

Interferometry with spin-squeezed electron states can be similarly described, by applying a similar interferometer sequence to a squeezed initial state, as illustrated in Fig. 1c.

The collective spin state at every stage can be visualized using the Wigner function [55] defined on the Bloch sphere, shown in Fig. 1a, c. All the expectation values and uncertainties of collective spin operators $\hat{S}_i$ can be calculated as expectation values using the Wigner function as a quasi-probability representation of the collective spin variables $(S_x, S_y, S_z)$.

## Quantifying spin squeezing

The accuracy of the interferometric phase measurements described above, with the phase determined according to Eq. (1), is limited by quantum projection noise. The root-mean-square phase uncertainty $\Delta\phi_{\mathrm{W}}$ can be calculated in terms of the spin operators [56]:

$$\Delta\phi_{\mathrm{W}} = \frac{\sqrt{\langle(\hat{S}_y - \langle\hat{S}_y\rangle)^2\rangle}}{|\langle\hat{S}_x\rangle|}. \tag{2}$$

The spin components here refer to the section of the MZI before the interaction with the sample, where the uncertainty ellipse is squeezed along the $y$-axis and expanded in the $z$-axis (Fig. 3c).

For a coherent spin state, the phase uncertainty of Eq. (2) follows the SQL, $\Delta\phi_{\mathrm{W}} = \Delta\phi_{\mathrm{SQL}} = N^{-1/2}$. For a general spin state, the phase uncertainty can be quantified by Wineland's spin-squeezing parameter $\xi^2$ [56–58], defined as $\xi^2 = (\Delta\phi_{\mathrm{W}})^2/(\Delta\phi_{\mathrm{SQL}})^2$. A spin-squeezed state is characterized by $\xi^2 < 1$, indicating that the phase uncertainty is reduced below the SQL.

An alternative measure of phase estimation accuracy is the quantum Fisher information $F$ [17,59], which quantifies the minimal possible phase measurement uncertainty for a given quantum state of electrons in the interferometer. For a pure state, the quantum Fisher information is given by $F = 4\langle(\hat{S}_z - \langle\hat{S}_z\rangle)^2\rangle$, and the related phase sensitivity is:

$$\Delta\phi_{\mathrm{F}} = F^{-1/2} = \frac{1}{2\sqrt{\langle(\hat{S}_z - \langle\hat{S}_z\rangle)^2\rangle}}. \tag{3}$$

In a conventional interferometer limited by the SQL, $F = N$, and in the Heisenberg limit, $F = N^2$. For a squeezed state, the increased uncertainty of $\hat{S}_z$ corresponds to increased Fisher information and reduced phase uncertainty $\Delta\phi_{\mathrm{F}}$.

Both the spin squeezing parameter $\xi^2$ and the quantum Fisher information $F$ are useful for quantifying the phase estimation accuracy. The quantum Fisher information-derived phase uncertainty $\Delta\phi_{\mathrm{F}}$ provides the theoretical lower bound of the phase uncertainty (the quantum Cramér Rao bound [60]), although a measurement protocol that can achieve this limit is only known in certain specific cases. In contrast, $\Delta\phi_{\mathrm{W}}$ indicates the phase uncertainty achievable using conventional interferometric protocols where the phase is determined according to Eq. (1). Notably, the two uncertainties coincide for a coherent spin state, retrieving $\Delta\phi_{\mathrm{SQL}}$.

We propose two methods for achieving electron spin squeezing. The first is interaction-based squeezing, which results in one-axis twisting. The second is measurement-based squeezing, which utilizes quantum non-demolition measurements. Both methods can potentially reach quantum Fisher information approaching the Heisenberg scaling. We further show that both methods can generate squeezed states ($\xi^2 < 1$) and therefore offer a practical path toward improving phase sensitivity beyond the shot-noise limit.

## Interaction-based squeezer

The proposed interaction-based squeezer utilizes the Coulomb repulsion between electrons to generate squeezing. Coulomb interactions at the electron densities typical for cryo-EM are not strong enough to significantly affect the electron's trajectories. However, these interactions can induce a significant phase shift in the electrons' wave functions. For instance, two electrons separated by 2.6 cm – a typical distance at an electron energy of 100 keV and a beam current of 1 nA (readily achievable with a field-emission electron gun) – experience an interaction-induced phase shift of approximately 0.51 radians after propagating for 1 meter.

Coulomb-mediated spin squeezing was introduced theoretically in [18]. In that work, the electron wave functions are assumed to extend over the entire interaction region. In contrast, TEM typically operates in a regime of well-separated electrons: the electron wave packet length is on the order of one micrometer [61,62], while their longitudinal separations are a few centimeters at typical TEM currents.

In this setting, a major challenge in utilizing the Coulomb interaction between electrons for spin squeezing is the stochastic distribution of their separations, which leads to significant phase noise. One possible way to mitigate this phase noise is to use the counter-propagating geometry considered in [18]. However, generating electrons in a coherent superposition of high-energy counter-propagating wave packets is a separate challenge that remains unsolved.

We propose a solution to the phase noise problem by replacing the pairwise, distance-dependent Coulomb interaction with a distance-independent, collective interaction. We envision passing the two arms of the interferometer through conductive channels (Fig. 2a). The channel length $l$ is assumed to be large enough for all $N$ electrons to be inside the channel simultaneously. The channel radius $r$ is designed to be much smaller than the distance between electrons in the same channel. The channels are also assumed to be either at low enough temperature or superconductive, so that the thermally induced dephasing [63] can be neglected.

In this setup, the Coulomb interaction will cause a phase shift in the interferometer that depends on the number of electrons in each arm. Specifically, the pair of conducting channels functions as a capacitor with mutual capacitance $C$, where the imbalance in electron numbers between the left and right arms induces an electric potential difference $\mathrm{V} = e(n_R - n_L)/(2C)$. This potential induces an interferometric phase shift in each electron, resulting in a collective electron-electron interaction. At the same time, the pairwise distance-dependent Coulomb interactions of electrons in the same channel are suppressed by the screening effect (SM Section 1.1) of the narrow conducting channels. Thus, passing the interferometer arms through the pair of conductive, capacitively coupled channels replaces the pairwise Coulomb interactions between electrons with a phase shift that depends only on the electron count in each arm. Here, we focus on the non-relativistic approximation, leaving a rigorous relativistic analysis to future work.

The channel response is assumed to be purely dispersive, so that the conducting channels return to their initial state after the electrons pass through, with no real excitations generated. Surface-plasmon excitation, the principal inelastic excitation channel, is exponentially suppressed beyond the evanescent range of the plasmon field [64], which is only tens of nanometers for the electron energies considered here. Experiments likewise show that conductor-induced decoherence decreases rapidly with electron–surface distance and is substantially weaker near highly conductive gold than near resistive silicon [65,66]. The channels can therefore be engineered with sufficiently large radii and low-loss conducting materials to suppress real excitations and realize the lossless regime assumed here.

The dependence of each electron's phase on the distribution of other electrons between the two channels creates quantum entanglement between the electrons. This entanglement can be described by including in the system Hamiltonian the total electrostatic energy of the capacitor $U = \frac{e^2(n_R - n_L)^2}{8\,C}$ (SM Section 1.1). Thus, the transformation performed by the two-channel device on the input electron state can be described by the operator:

$$S_{\text{int}} = e^{-\frac{i\chi_{\text{int}}\hat{S}_z^2}{2}}. \tag{4}$$

with the interaction strength parameter $\chi_{\text{int}} = \frac{e^2 l}{v\hbar C}$, with $l$ being the channel length and $v$ the electrons' velocity. The transformation produced by this operator is known as one-axis twisting, and it generates a spin squeezed state that can be used for sub-shot-noise metrology as illustrated in Fig. 2.

For example, for cylindrical channels, shown in Fig. 2a., the mutual capacitance can be approximated as $C = \frac{l\varepsilon_0\pi}{\text{arccosh}\left(\frac{d}{2r}\right)}$ [67]. In this case, the interaction strength takes the form $\chi_{\text{int}} = 4\frac{\alpha}{\beta}\text{arccosh}\left(\frac{d}{2r}\right)$, where $\alpha$ is the fine structure constant and $\beta = \frac{v}{c}$. The resulting $\chi_{\text{int}}$ is independent of $l$ because the mutual capacitance of the channels is proportional to their length. The interaction strength can be controlled by setting the geometry of the channels and the electron kinetic energy, $E = m_e c^2\left(\frac{1}{\sqrt{1-\beta^2}} - 1\right)$, where $m_e$ is the electron's mass. For instance, for $d = 10r$ and electron kinetic energy of $E = 100$ keV, the interaction strength is $\chi_{\text{int}} = 0.122$, which for $N = 40$ electrons provides more than a factor of 2 improvement in the phase uncertainty, from $\Delta\phi_{\text{SQL}} \approx 0.158$ to $\Delta\phi_{\text{W}} \approx 0.063$. The Coulomb squeezer can be adjusted to larger $N$ values for which the optimal $\chi_{\text{int}}$ is smaller, e.g., by reducing the distance $d$ between the channels.

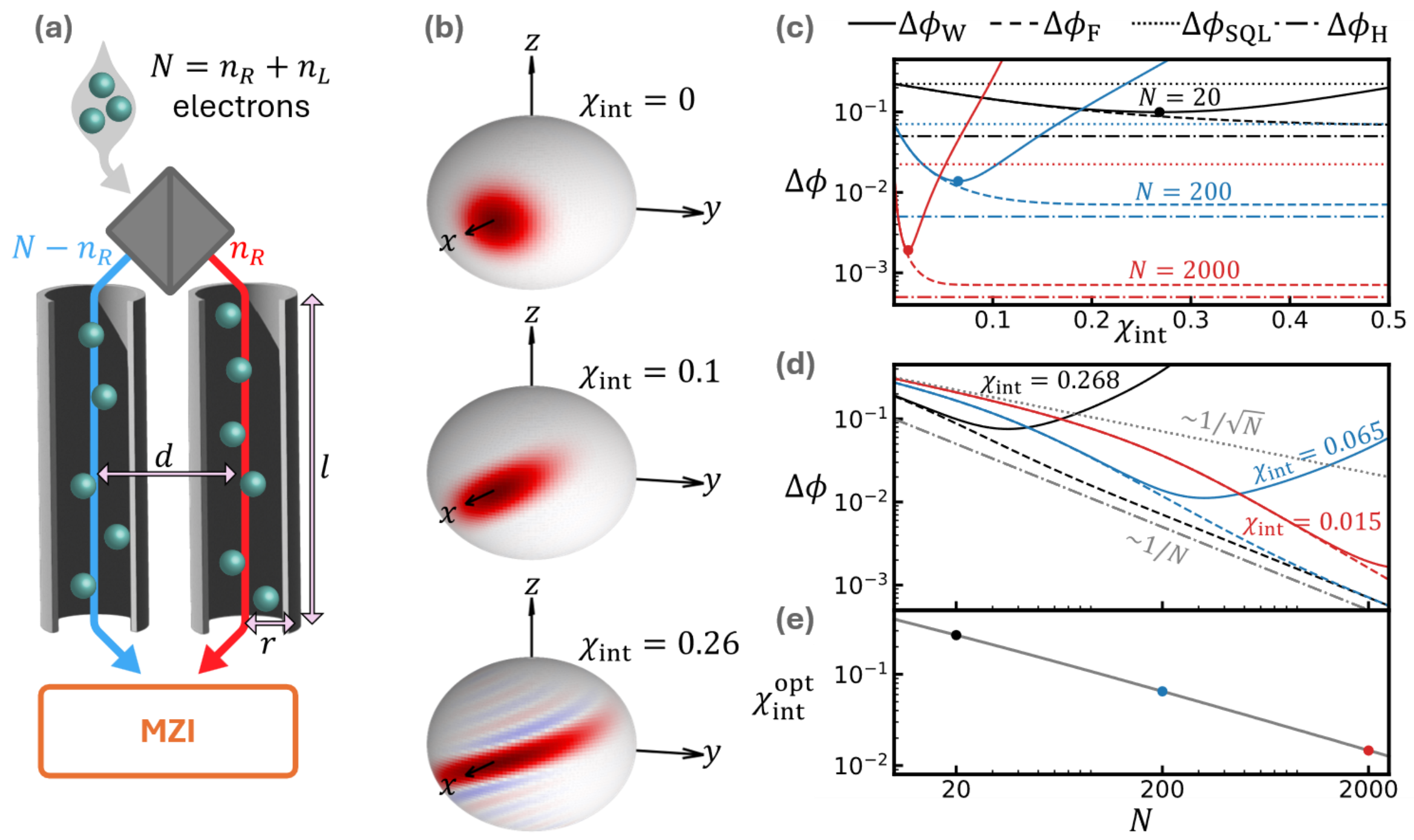

**Fig. 2: Interaction-based squeezer. Free-electron analogue of one-axis twisting. (a) Squeezer schematic:** After the beamsplitter, the electrons are passed through two conductive channels, inducing entanglement through capacitive coupling. **(b) Wigner functions of the resulting squeezed states** correspond to $\chi_{\rm int} = 0, 0.1, 0.26$ and $N = 20$ electrons. A subsequent rotation (around the $x$-axis) is necessary to align the squeezed axis properly for improving phase sensitivity. **(c) Phase uncertainty $\Delta\phi$ as a function of $\chi_{\rm int}$**, comparing: $\Delta\phi_{\rm W}$ (solid lines) the phase uncertainty of the squeezed interferometer with highlighted points marking minimal values, $\Delta\phi_{\rm F}$ (dashed lines) the best achievable phase uncertainty, $\Delta\phi_{\rm SQL}$ (dotted lines) the standard quantum limit, and $\Delta\phi_{\rm H}$ (dashdotted lines) the Heisenberg limit. Different colors correspond to $N = 20, 200, 2000$ electrons. **(d) Phase uncertainty as a function of $N$** with different colors correspond to $\chi_{\rm int} = 0.268, 0.065, 0.015$ shown along $\Delta\phi_{\rm SQL}$ ($\chi_{\rm int} = 0$) (dotted lines) and $\Delta\phi_{\rm H}$ (dashdotted lines). **(e) Optimal interaction strength values $\chi_{\rm int}$** that achieve the minimal phase uncertainty $\Delta\phi_{\rm W}$ as a function of $N$. The highlighted points correspond to the minimal phase uncertainties highlighted in (c), for $N = 20, 200, 2000$.

The one-axis twisting operator generated by the Coulomb squeezer produces a state with a minimal spin projection uncertainty along an oblique direction, as shown in Fig. 2b. This tilt can be compensated by rotating the collective state by an extra angle $\delta = \frac{1}{2}\arctan\left(\frac{4\sin(\chi/2)\cos^{N-2}(\chi/2)}{1-\cos^{N-2}(\chi)}\right)$ [36,45], which can be implemented by adjusting the coupling coefficient of the input beamsplitter of the MZI.

Performing interferometric measurements on input states prepared via one-axis twisting provides a substantial quantum enhancement that grows with $N$. The phase uncertainties $\Delta\phi_{\rm W}$ (Eq. (2)) and $\Delta\phi_{\rm F}$ (Eq. (3)) as a function of $N$ and $\chi_{\rm int}$ are plotted in Fig. 2c and Fig. 2d. For any given $N$, the phase uncertainty improves with increasing interaction strength $\chi_{\rm int}$ up to an optimal squeezing strength $\chi_{\rm int}^{\rm opt}$. Fig. 2e shows the optimal squeezing strength as a function of $N$, calculated by finding the minimum value of $\Delta\phi_{\mathrm{W}}$. At the optimal interaction strength, the spin squeezing parameter scales as $\xi^2 \propto N^{-2/3}$ , which corresponds to the best achievable squeezing for one-axis twisting [45,58]. Thus, the capacitive Coulomb squeezer can provide phase measurement accuracy well beyond the SQL, although not quite reaching the Heisenberg limit.

Interestingly, conventional interferometric measurements with one-axis-twisted states do not reach the Heisenberg limit because such measurements do not fully utilize the information contained in the output states, as can be assessed using Fisher information. The lower bound of the phase uncertainty for the squeezed state scales as $\Delta\phi_{\mathrm{F}} \propto N^{-1}$ (Fig. 2d), suggesting the potential to reach the Heisenberg limit. Indeed, at $\chi_{\mathrm{int}} = \pi$ the interaction-based squeezer generates a GHZ state [68], which gives the minimal phase uncertainty of $\Delta\phi_{\mathrm{F}} = N^{-1}$ [69]. In this case, the Heisenberg limited accuracy can be realized by performing a parity measurement on the output state of the interferometer [70]. Experimental realization of this scheme, however, is more difficult than spin-squeezed interferometry, as it is more sensitive to electron loss and dephasing in the sample and puts more stringent requirements on electron detection accuracy.

## Measurement-based squeezer

In a measurement-based squeezer, measuring the number of electrons in one (or both) of the interferometer arms reduces the uncertainty of the number of electrons passing through that arm, thereby creating a squeezed state in $\hat{S}_z$ (Fig. 3). For the two electron paths to remain coherent after measurement, the measuring device must be insensitive to any properties of the electron's state, such as energy and arrival time, which may distinguish between individual electrons. In other words, the measurement must determine the total number of electrons in each arm of the interferometer without measuring the which-path information of individual electrons. This method, illustrated in Fig. 3a, is inspired by the use of quantum non-demolition measurements, which have been explored for enhanced atomic interferometry [18,58,71,72].

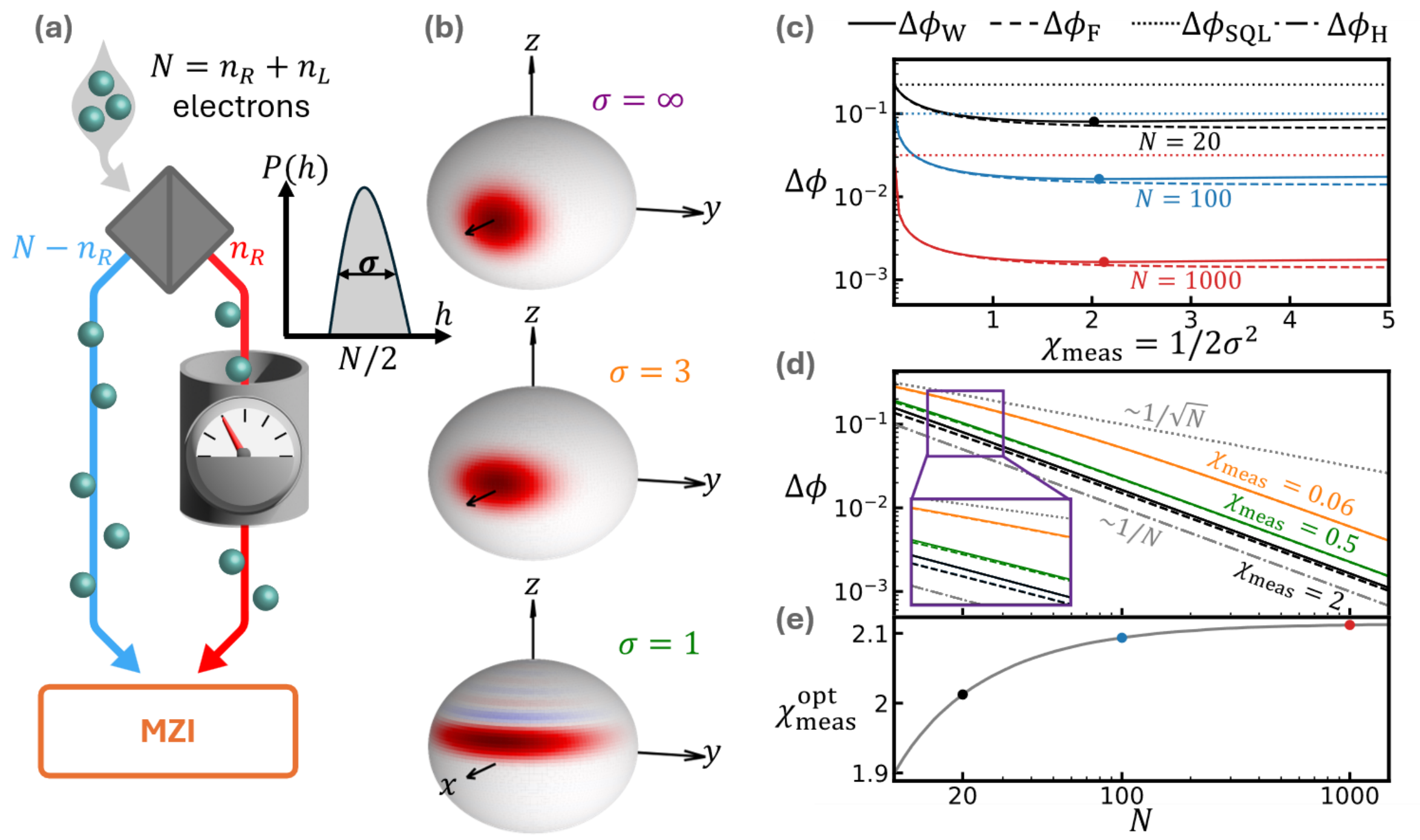

**Fig. 3: Measurement-based squeezer. Free-electron analogue of quantum non-demolition measurement. (a) Squeezer schematic:** After the beamsplitter, the right path passes through the measurement device, which counts the number of electrons. **(b) Wigner functions of the resulting squeezed states** correspond to $\sigma = \infty, 3, 1$, ($\chi_{meas} = 0, 0.06, 0.5$) and $N = 20$ electrons, with measurement result $h = 7$. **(c) Phase uncertainty $\Delta\phi$ as a function of $\chi_{meas}$**, comparing: $\Delta\phi_W$ (solid lines) the phase uncertainty of the squeezed interferometer, with highlighted points marking minimal values, $\Delta\phi_F$ (dashed lines) the best achievable phase uncertainty, $\Delta\phi_{SQL}$ (dotted lines) the standard quantum limit, and $\Delta\phi_H$ (dashdotted lines) the Heisenberg limit. Different colors correspond to $N = 20, 100, 1000$ electrons. **(d) Phase uncertainty as a function of $N$** with different colors correspond to $\chi_{meas} = 0.06, 0.5, 2$ ($\sigma = 3, 1, 0.5$) shown along $\Delta\phi_{SQL}$ ($\chi_{meas} = 0$) (dotted lines) and $\Delta\phi_H$ (dashdotted lines). **(e) Optimal interaction strength values $\chi_{meas}$** that achieve the minimal phase uncertainty $\Delta\phi_W$ as a function of $N$. The highlighted points correspond to the minimal phase uncertainties highlighted in (c), for $N = 20,\ 100, 1000$.

The quantum non-demolition measurements can be described in terms of Kraus operators [73]. We analyze here the case where the measurement noise is normally distributed around zero with variance $\sigma^2$. In the case of, for example, measuring the number of electrons in the right arm $\hat{n}_R$, the Kraus operator is:

$$\hat{K}_{meas}(h) = \frac{1}{\text{norm}} e^{-\frac{\chi_{meas}}{2}(\hat{n}_R - h)^2} = \frac{1}{\text{norm}} e^{-\frac{\chi_{meas}}{2}\left(\frac{N}{2} - \hat{S}_z - h\right)^2}, \quad (5)$$

where $h$ is the measurement result and $\chi_{meas} = \frac{1}{2\sigma^2}$. The measurement will produce different results on every run, due to both the quantum uncertainty of $\hat{n}_R$ and the measurement error $\sigma$. For any $h$, the Kraus operator action on the electron state $|\psi_0\rangle$ produces a state $\hat{K}_{meas}(h)|\psi_0\rangle$, squeezed in the $z$ direction as well as offset from the equator of the Bloch sphere by $\frac{N}{2} - h$ (Fig.

3b). The sub-shot-noise limited phase estimate is then obtained by subtracting $h$ from the final measurement of $\hat{S}_z$.

Performing interferometric measurements on input states prepared via quantum non-demolition measurements provides a substantial quantum enhancement that grows with $N$. To quantify it, for each measurement result $h$ we calculate the spin-squeezing parameter $\xi^2(h)$ and the quantum Fisher information $F(h)$. We then average over all possible measurement result $h$ with the probabilities $p(h) = \left|\hat{K}_{\text{meas}}(h)|\psi_0\rangle\right|^2$ to obtain the phase uncertainties: $\Delta\phi_{\text{W}} = \frac{1}{\sqrt{N}}\sqrt{\int_{-\infty}^{+\infty} p(h)\xi^2(h)dh}$ and $\Delta\phi_F = \int_{-\infty}^{+\infty} \frac{p(h)}{\sqrt{F(h)}}dh$. We find analytically that $\Delta\phi_{\text{F}} = 1/\sqrt{N + \frac{N^2-N}{2}(1 - e^{-\chi_{\text{meas}}})}$ (SM Section 2.4), which for small enough $\sigma$, follows Heisenberg scaling.

The phase uncertainties $\Delta\phi_{\text{W}}$ (calculated numerically) and $\Delta\phi_{\text{F}}$ as a function of $N$ and $\chi_{\text{meas}}$ are plotted in Fig. 3c and Fig. 3d. For any given $N$, the phase uncertainty improves with increasing $\chi_{\text{meas}}$ up to an optimal squeezing strength $\chi_{\text{meas}}^{\text{opt}}$. Fig. 3e shows the optimal squeezing strength as a function of $N$, calculated by finding the minimum value of $\Delta\phi_{\text{W}}$.

Remarkably, Fig. 3e shows that the optimal squeezing strength is in the vicinity of $\chi_{\text{meas}}^{\text{opt}} = 2$, corresponding to $\sigma = \frac{1}{2}$, for any number of electrons $N$. Intuitively, this can be understood as follows. To provide a metrological improvement, $\sigma$ must be smaller than the quantum projection noise: $\sigma < \sqrt{N/2}$. In the limit of precise electron-number measurement where $\sigma \ll \frac{1}{2}$, each measurement projects the electrons into a Dicke state, which cannot be readily used for interferometry. In contrast, in the weak squeezing regime where $1/2 \ll \sigma < \sqrt{N/2}$, phase uncertainty decreases with decreasing $\sigma$. Thus, useful squeezing can be achieved in the intermediate range where $1/2 \lesssim \sigma \ll \sqrt{N}$, with the optimal squeezing strength $\sigma \approx 1/2$.

As an example, measurement-based free-electron spin squeezing can be implemented using electron interaction with a microwave cavity [74]. Following the first beamsplitter shown in Fig. 3a, one of the interferometer arms is passed through a microwave cavity designed so that only one of its modes couples to the electrons, e.g., via phase-matching [75–78]. The electron beam must be temporally modulated into one or more bunches synchronized with the phase of the cavity mode [79], up to an integer multiplication of its oscillation cycle. The electrons passing through the cavity collectively create a photonic coherent state with amplitude $\alpha = n_R\sqrt{\chi_{\mathrm{meas}}/2}$. Here, the interaction strength $\chi_{\mathrm{meas}}$ equals the parameter $2|g_Q^2|$ that is often used to characterize the coupling between the electrons and the cavity [80–85]. The squeezing is then created by a measurement of the amplitude $\alpha$ [86].

Strong interaction between the cavity and the electrons is needed to achieve strong spin squeezing. This condition can be relaxed by initializing the cavity with a photonic squeezed vacuum state (see SM Section 2.1 for detailed results and derivation). Thermal cavity excitation and imperfect quadrature readout increase the effective measurement noise and therefore reduce the attainable measurement strength (SM Section 3.3). Preserving coherence between the interferometer arms additionally requires the cavity to measure only the total electron number in the interacting arm, without resolving individual electron arrival times or other distinguishing information. The cavity should have a sufficiently long lifetime to retain the collective displacement produced by the entire electron bunch (or multiple thereof). A subsequent quadrature readout measures the cavity state. In addition, the electron energy uncertainty must be sufficiently large that the energy exchanged with the cavity does not reveal which interferometer arm was taken.

# Discussion

Spin-squeezing electron interferometry can provide a substantial improvement of the phase measurement accuracy at the electron doses suitable for cryo-EM. We note that even a moderate improvement in the SNR of an imaging system can have a profound effect on the field, as illustrated by the transition from scintillator-based cameras to direct electron detectors, which increased the SNR roughly twofold, playing a major role in bringing about the “resolution revolution” in cryo-EM.

Electron beam-induced damage constrains the dose to about 20 electrons per square angstrom [1], or approximately 1.5x higher at liquid Helium temperatures [87]. Consequently, suitable batch sizes range from about $N \approx 40$ for a typical high-resolution cryo-EM pixel size of 1 Å, to $N \approx 2000$ for a resolution of 10 Å, suitable for cryo-EM of larger structures. Note that on average only ½ $N$ electrons pass through the sample at each pixel in this scheme.

For example, at $N = 40$ electrons, an optimal interaction-based squeezer will require $\chi_{\mathrm{int}} \approx 0.177$, providing more than a factor of 2 improvement in the phase uncertainty, from $\Delta\phi_{\mathrm{SQL}} \approx 0.158$ to $\Delta\phi_{\mathrm{W}} \approx 0.055$. However, even with suboptimal squeezing of $\chi_{\mathrm{int}} \approx 0.122$ that can be achieved using capacitors with an aspect ratio of $d = 10r$, the phase uncertainty also improves by more than a factor of 2 to $\Delta\phi_{\mathrm{W}} \approx 0.063$. For the same electron dose, an optimal measurement-based squeezer will require $\chi_{\mathrm{meas}} \approx 2.06$, providing a factor 4 improvement in phase sensitivity of $\Delta\phi_{\mathrm{W}} \approx 0.04$.

As another example, at $N = 2000$, optimal squeezers with $\chi_{\mathrm{int}} \approx 0.015$ and $\chi_{\mathrm{meas}} \approx 2.12$, the phase uncertainty improves by more than an order of magnitude, from $\Delta\phi_{\mathrm{SQL}} \approx 0.023$ to $\Delta\phi_{\mathrm{W}} \approx 0.002$ and $\Delta\phi_{\mathrm{W}} \approx 0.0008$, for the interaction-based and measurement-based squeezers, respectively. Importantly, even with suboptimal squeezing of, for example, $\chi_{\mathrm{meas}} = 0.5$, the phase uncertainty still improves by more than an order of magnitude, from $\Delta\phi_{\mathrm{SQL}} \approx 0.023$ to $\Delta\phi_{\mathrm{W}} \approx 0.0011$. A factor-of-two improvement in phase uncertainty only requires $\chi_{\mathrm{meas}} \approx 0.003$, corresponding to coupling of only $g_{\mathrm{Q}} \approx 0.04$.

These results demonstrate that optimal squeezing strength can be achieved with realistic squeezer parameters across the entire resolution range relevant to cryo-EM, establishing spin squeezing as a natural modality of quantum metrology for application in electron microscopy.

In the initial analysis provided in this paper, we did not consider electron losses to both elastic and inelastic scattering in the sample, and possible losses in the elements of the interferometer. Spin-squeezing is known to be among the most robust quantum-metrology approaches, less sensitive to electron losses than other entanglement-based quantum metrology schemes. The mean free path of electrons in amorphous ice ranges from roughly 300 nm at 300 keV to 200 nm at 100 keV [88]. Consequently, in a typical thin cryo-EM sample of about 30 nm, the share of electrons lost to inelastic scattering is in the range of 0.1 to 0.2. For thicker samples, such as 100-200 nm lamellae used in cryo-ET [89], imaged at 300 keV, that share is in the range of 0.3 to 0.5. While this level of electron loss limits the useful strength of squeezing [17], it still allows for a significant metrological improvement. This relatively robust behavior of spin-squeezed electron states contrasts with more fragile fully entangled states such as GHZ, which loses the phase information with loss of even a single electron in a batch.

Losses can occur at each stage of our protocol: in the sample, beamsplitters, conducting channels or cavity, and electron detector. SM Section 3.1 considers a qualitative loss model for their combined effect as independent Bernoulli transmission with total transmission $T$ and interferometric visibility $V$, which results in the conventional-readout phase uncertainty approximately $\left(\Delta\phi_{\mathrm{W,loss}}\right)^2 \approx [\xi^2 + (1-T)/T]/(NV^2)$. This expression separates the intrinsic squeezing $\xi^2$ from the binomial noise added by loss or missed detection. The minimum $T$ required to obtain a net gain over the SQL requires $\xi^2 + (1-T)/T < V^2$, or equivalently $T > 1/(1 + V^2 - \xi^2)$. For $V \approx 1$ and the squeezing levels discussed above, the required transmission is $T > 1/2$ for conventional interferometric readout.

A general analysis of the fundamental quantum limit including losses should consider the quantum Fisher information for a mixed state (SM Section 3.2). For fixed nonzero loss, the Heisenberg scaling of highly entangled states is not generally retained asymptotically, and this limitation is particularly severe for NOON/GHZ-type states. However, this does not imply that the quantum advantage disappears: the dose-limited analysis of [17] shows information-per-dose advantage for any nonzero sample transmissivity. Increasing sample loss continuously reduces this advantage, which vanishes only in the limit of complete loss of electrons.

In addition to the squeezer devices discussed above, experimental realization of a spin-squeezed electron Mach-Zehnder interferometer requires a coherent, low-loss beamsplitter to split and recombine the electron waves, and a low-noise electron detector. We now briefly review the state of technology for each of those components.

Electron beamsplitters: Conventionally, electron interference has been observed by electron holography using an electrostatic biprism (a "wavefront splitter" device) and observing electron interference fringes. However, to implement spin squeezing interferometry schemes envisioned here, a true "amplitude splitter" is needed. Interferometry with amplitude-splitting electron beamsplitters have been realized with membrane-based diffraction gratings [90–93,15], although such membranes are inherently lossy due to electron scattering. A theoretical proposal has been made to create a nearly lossless beamsplitter based on a diffraction grating embedded in an electrostatic electron resonator [94]. In parallel, significant research efforts are dedicated to the development of a practically lossless electron beamsplitters based on electron confinement in guiding channels by either RF ponderomotive potential [95,96] or auto-ponderomotive potential created by electrostatic fields of alternating polarity [97,98]. These on-going efforts make it probable that a nearly lossless beamsplitter will be created over the next few years.

Electron detectors: Electron detectors have been developing rapidly in recent years and have already reached a sufficient signal-to-noise ratio to enable squeezed interferometry.

Several types of detectors (monolithic CMOS direct electron detectors, hybrid pixel sensors) have been developed that operate in the electron-counting mode, detecting individual electron arrival events. Quantum efficiency of such detectors can be as high as 0.94 [99] or even 0.98 [100]. Thus, already existing electron detectors technology can be used to read out the number of electrons in each port of the interferometer with deeply sub-shot-noise limited accuracy.

Measurement-based free-electron spin squeezing can be implemented using an optical structure rather than the microwave cavity considered above. An optical implementation can rely on efficient cathodoluminescence [101–103] whereby the electron becomes entangled with the photon it emits [104,105]. Sufficient energy uncertainty to maintain path interference after the interaction can be achieved using short (e.g., attosecond [106–108]) electron pulse trains. As another alternative, free-space electrons can be entangled with ancillary quantum systems such as localized spins [109], trapped-ion quantum information processors [110], or superconducting qubits [11,19,20], all of which could serve as the measurement interface in the implementation of our proposed measurement-based squeezer.

The interaction- and measurement-based squeezing mechanisms can also be combined. For example, weak one-axis twisting could first generate number–phase correlations, followed by a QND measurement that further conditions the collective state. Such a hybrid protocol introduces a tradeoff between the interaction and measurement strengths and could reduce the strength required from either individual squeezing stage. Determining whether this also improves the optimal sensitivity or loss tolerance requires a joint optimization over the two squeezing strengths and experimental noise parameters, which we leave for future work (see SM Section 4).

The concept of free-electron spin-squeezing can thus contribute to the wider field of free-electron quantum optics [32,111], which so far mostly focused on electron-light interactions [86,112–116], rather than on electron phase sensitivity. The approach suggested here for measurement-based electron spin squeezing can help create various forms of entangled free-electron and photonic states that can be harnessed in other areas of quantum science [82,83,86,104,105,115,117].

We analyzed electron spin squeezing in a scanning interferometer configuration that surveys samples pixel by pixel. Signal acquisition can be parallelized with a multi-channel squeezer of the same conceptual design. Alternatively, a wide-field spin-squeezed imaging can potentially be implemented using a squeezer device capable of multi-mode entanglement, based, e.g., on collective emission of light by electrons interacting with an extended object, such as a membrane or a sheet of cathodoluminescent material.

In summary, we have shown that the concept of spin squeezing can be adapted to transmission electron microscopy to overcome the seemingly fundamental shot-noise limit on the SNR. The necessary pseudospin-squeezed electron states can be generated in a variety of ways with existing and emerging technologies, providing a robust and practical pathway toward quantum-enhanced electron microscopy. By contributing to the development of cryo-EM and cryo-ET, spin-squeezed electron microscopy may become one of the few applications of quantum metrology whose significance extends beyond the field of quantum physics.

# Methods

## Computational representation

The calculations were performed in Python using NumPy and SciPy for numerical arrays and combinatorial coefficients, QuTiP for collective-spin operators and quantum-state calculations, and Matplotlib for visualization. Because all transformations considered here preserve permutation symmetry between the path pseudospins, the $N$-electron state was represented in the symmetric Dicke basis of dimension $N+1$, rather than in the full $2^N$ dimensional tensor-product Hilbert space. All calculations were deterministic and involved no stochastic sampling or fitting of the numerical curves. Reported optimal parameters correspond to the minimum value on the specified numerical grid, without interpolation between adjacent grid points.

## Interaction-based simulations

For the interaction-based squeezing calculations, the one-axis-twisting operator $S_{\text{int}}$ of Eq. (4) was applied to the initial coherent spin state, followed by additional rotation that aligns the squeezed quadrature with the interferometric measurement axis. Collective-spin expectation values and variances were calculated from the resulting state and used to evaluate the corresponding phase uncertainties $\Delta\phi_{\text{W}}$ (Eq. (2)) and $\Delta\phi_{\text{F}}$ (Eq. (3)). Unitary collective-spin rotations and the one-axis-twisting operator were applied by matrix exponentiation. For the scans in Fig. 2c, $\chi_{\text{int}}$ was sampled uniformly between 0 and 0.5 using 5,000 grid points. The curves in Fig. 2d were calculated at the fixed $\chi_{\text{int}}$. For each value of $N$ in Fig. 2e, the optimal interaction strength was determined by minimizing the Wineland phase uncertainty.

### Measurement-based simulations

For measurement-based squeezing, the Kraus operator of Eq. (5) was applied to the initial coherent spin state for each possible continuous measurement result $h$. Conditional states produced by the measurement Kraus operator were normalized before expectation values and variances were evaluated and used to evaluate the corresponding phase uncertainties $\Delta\phi_{\text{W}}$ and $\Delta\phi_{\text{F}}$ averaged over all possible measurements, as in SM Section 2.6. The outcome integral was evaluated on a uniform grid extending from $-10\sigma$ to $N + 10\sigma$. At least 1001 outcome points were used, with the number increased to $4N + 1$ for larger ensembles. The curves in Fig. 3c were calculated for $N = 20, 100$, and $1000$. The fixed-strength curves in Fig. 3d used $\chi_{\text{meas}} = 0.06$, $0.5$, and $2$, corresponding approximately to $\sigma = 3$, $1$, and $0.5$, respectively. For each $N$ in Fig. 3e, the optimal measurement strength was found by minimizing the outcome-averaged Wineland phase uncertainty in the vicinity of $\chi_{\text{meas}} = 2$.

### Wigner-function visualization

The collective-spin Wigner functions were calculated using QuTiP. The functions were evaluated on a spherical grid containing 80 polar-angle points and 160 azimuthal-angle points and were displayed on the unit Bloch sphere using a diverging color scale centered at zero. The interaction-based states in Fig. 2b were calculated for $N = 20$ and $\chi_{\text{int}} = 0$, $0.1$, and $0.26$, including the quadrature-alignment rotation. The measurement-based states in Fig. 3b were calculated for $N = 20$, measurement outcome $h = 7$, and $\sigma = \infty$, $3$, and $1$.

## Declaration statements

### Data Availability

All data supporting the findings of this study are available within the paper and its Supplementary Materials. The numerical data underlying the main and supplementary figures and supporting the findings of this study can be generated using the code available in the public GitHub repository cited in the Code Availability statement.

## Code Availability

The Python/QuTiP code used to generate the numerical results and figures is publicly available at: https://github.com/ShiranEv/electron-spin-squeezing.

## Acknowledgments

This research is funded by the European Union- ERC COG, QinPINEM, Project Number 101125662. This research is also funded (in part) by the Gordon and Betty Moore Foundation, through Grant GBMF11473. O.S. acknowledges financial support from the Gordon and Betty Moore Foundation through Grant GBMF12217, the Israel Science Foundation (ISF), and the Center for New Scientists at the Weizmann Institute of Science. E.S. acknowledges financial support from the Israel Science Foundation (ISF), the Directorate for Defense Research and Development (DDR&D), the Minerva Foundation with funding from the Federal German Ministry for Education and Research, and the Center for New Scientists at the Weizmann Institute of Science. S.E.-H. acknowledges financial support from the Ariane de Rothschild Women Doctoral Program. R.R. acknowledges financial support from the ADAMS Fellowship Program of the Israel Academy of Sciences and Humanities.

## Author Contributions

S.E.-H. and E.N. contributed equally to this work and developed the theoretical framework, performed the analysis, and wrote the initial draft of the manuscript. R.B.-M., A.G., and R.R. contributed to the theoretical development, calculations, and interpretation of the results. E.S. contributed to conceptual development and theoretical analysis. O.S. and I.K. supervised the project, contributed to the conceptual design, and guided the research. All authors discussed the results and contributed to the writing and revision of the manuscript.

## Competing Interests

The authors declare no competing financial or non-financial interests.